\documentclass[ aps, prl, amsmath,amssymb, reprint]{revtex4-2}
\makeatletter  \newcommand{\Rmnum}[1]{\expandafter\@slowromancap\romannumeral #1@} \makeatother
\usepackage[utf8]{inputenc} \usepackage[T1]{fontenc} \usepackage{CJK}
\usepackage{graphicx}  \usepackage{epstopdf} \usepackage{dcolumn} \usepackage{booktabs} \usepackage{array} \usepackage{float} \usepackage{tabularx} \usepackage{multirow}
\usepackage{newtxtext} \usepackage{newtxmath}
\usepackage{bm}
\usepackage{siunitx} \usepackage{xcolor} \usepackage{pifont} \usepackage[version=4]{mhchem} \graphicspath{{figs/}{figsgaoerb/}}
\usepackage{hyperref} \hypersetup{ colorlinks=true, urlcolor=blue, citecolor=blue, linkcolor=blue }
\usepackage{upgreek}

\begin{document}
	
	\title{Hydrogen Reorganization in Hot Dense Ammonia}	
	
	\author{Yu Tao}
	\affiliation{Institute of Atomic and Molecular Physics, Sichuan University, Chengdu 610065, China}
	
	\author{Jingyi Liu}
	\affiliation{Institute of Atomic and Molecular Physics, Sichuan University, Chengdu 610065, China}
	
	\author{Li Lei*}
	\affiliation{Institute of Atomic and Molecular Physics, Sichuan University, Chengdu 610065, China}

	\begin{abstract}
		Hydrogen is known to form stable compounds with a variety of simple molecular solids under high pressure. However, ammonia hydride has not been experimentally observed. Here, through a series of laser-heating diamond anvil cell (LHDAC) experiments, we observed that transparent ionic ammonia transformed into an opaque phase above 150 GPa and 1000 K.  Raman spectroscopy reveals a pronounced structural change, accompanied by the emergence of two distinct asymmetric H$_2$ vibrons ($\nu_1$ and $\nu_2$). A new ionic phase formed following thermally induced dissociation and hydrogen reorganization in hot dense ammonia. The combination of experimental results and  first-principles calculations identifies this phase as the theoretically predicted ionic $P4_12_12$ phase of NH$_7$ , which consists of NH$_4^+$, H$^-$and H$_2$ units. Our results suggest that the formation of ammonia hydrides follows a thermodynamic pathway distinct from those of other H$_2$-containing mixtures, revealing a new ionic transformation pathway driven by hydrogen reorganization in hot dense ammonia.
		
	\end{abstract}
	
	
\maketitle
	
	\textit{Introduction}---Hydrogen mixtures with simple molecular species are considered important constituents of planetary interiors\cite{1995_Nature_Loubeyre,1996_Science_Finger,2024_PRB_Yan,2010_NC_Somayazulu}. Currently, hydrogen has been experimentally identified to form stable van der Waals compounds with a variety of molecular solids, including Ar(H$_2$)$_2$, CH$_4$(H$_2$)$_2$, N$_2$(H$_2$)$_2$, and H$_2$O(H$_2$)$_2$ \cite{1994_PRL_Loubeyre,2022_PRL_Ranieri,2017_PNAS_Ji, 2014_NC_Spaulding, 2014_SC_Qian}. Theoretical studies have predicted potentially stable ammonia hydrides in the NH$_3$--H$_2$ mixture under high pressure, including $R\bar{3}m$ and $P4_12_12$ phase of NH$_7$, and $Pm$ phase of NH$_{10}$\cite{2019_JPCL_Song,2024_PRM_He}. However, although compression of NH$_3$--H$_2$ mixtures can enhance hydrogen bonding and promote proton-exchange reactions, NH$_3$ and H$_2$ remain immiscible over a broad pressure range \cite{2011_JPCA_Chidester,2014_JCP_Borstad}. Currently, no ammonia hydride formed through a direct reaction between NH$_3$ and H$_2$ has been experimentally identified.

Unlike most molecular hydrides, ammonia exhibits pronounced proton-transfer and ionization behavior under extreme conditions. Under static compression, solid ammonia has been experimentally shown to transform into ionic and superionic states \cite{2012_PRL_Ninet,2014_NC_Palasyuk,2026_JPCL_Tao,2026_PRB_Guigue,2021_PNAS_Kimura}. In the ionic phase, intermolecular proton transfer converts molecular NH$_3$ into ionic species (NH$_4^+$ NH$_2^-$). At elevated temperatures, rapid proton diffusion through hot dense ammonia gives rise to superionic behavior. The experimentally observed superionic ammonia (phase $\upalpha$) was obtained by laser heating the molecular phase V. At the same time, ammonia can undergo partial decomposition at high temperatures, accompanied by the formation of molecular hydrogen \cite{2012_JCP_Ojwang,2019_PRB_Ninet}. Nevertheless, no reaction between ammonia and the H$_2$ was observed, indicating that even under high-temperature conditions the molecular V phase does not readily form an ammonia hydride. These experimental behaviors contrast with theoretical predictions that stable NH$_3$--H$_2$ compounds may exist under compression even at room temperature\cite{2019_JPCL_Song}.

\begin{figure}[b]
	\centering
	\includegraphics[width=\columnwidth]{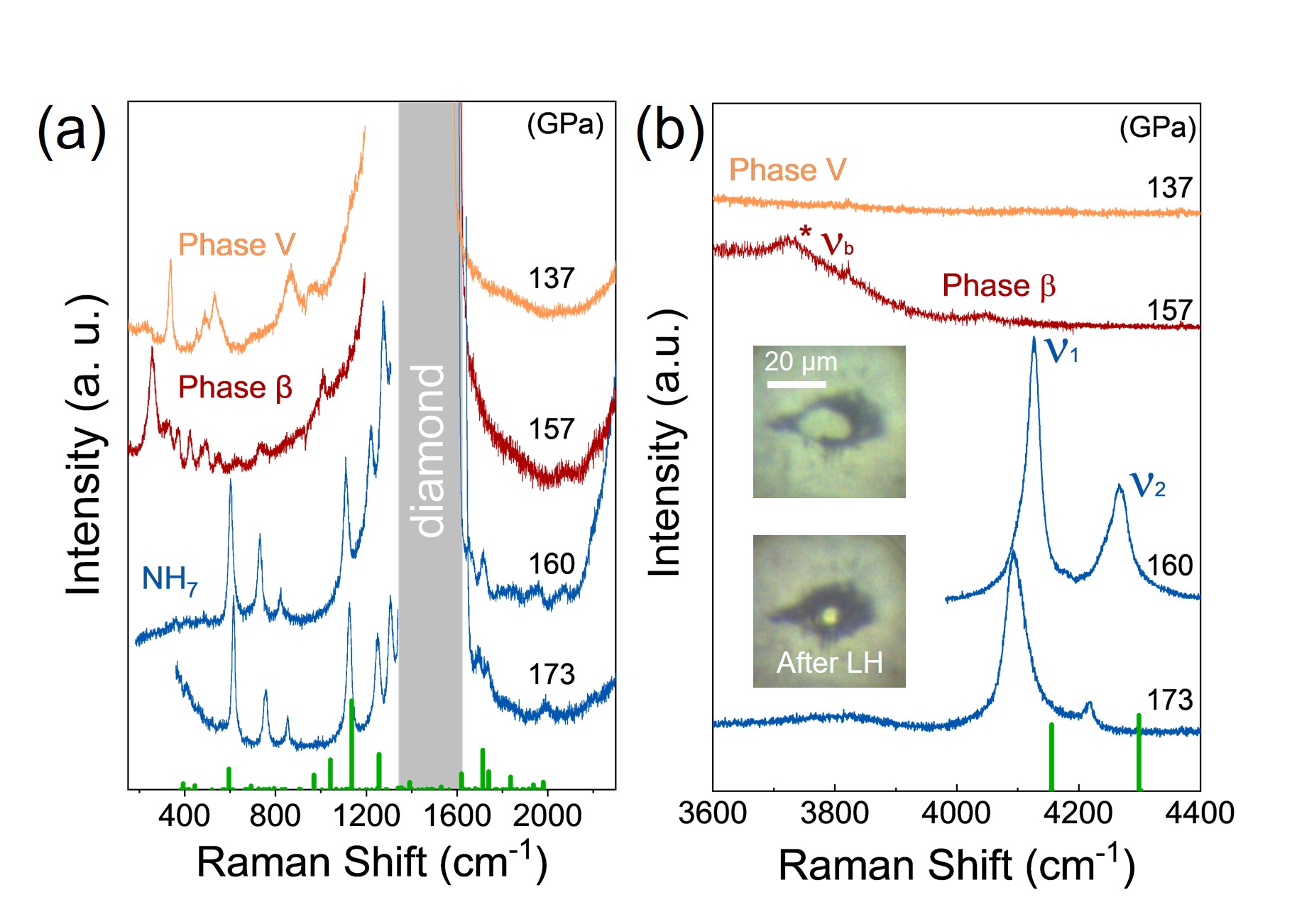}
	\caption{Representative Raman spectra of solid ammonia collected at different pressures before and after laser heating. (a) Raman spectra in the lattice-mode and N–H bending-mode regions. (b) Raman spectra in the stretching mode region. Phase V and phase $\upbeta$ are the molecular and ionic phase of solid ammonia, respectively, before laser heating. $\nu_{b}$ denotes the vibron mode of phase $\upbeta$.  $\nu_1$ and $\nu_2$ denote the H$_2$ vibron modes in NH$_7$. NH$_7$ are the ionic phase formed after heating. The green solid lines at the bottom show the DFT-calculated Raman spectra at 150 GPa. A global offset of 60 cm$^{-1}$ has been applied for clarity. The pressure dependences of the most intense Raman modes identified experimentally and by DFT calculations were further compared over 120--200 GPa (Fig. S11). The inset shows optical images of the sample chamber before and after laser heating. NH$_7$ was observed in the opaque region.}
	\label{fig:fig1}
\end{figure}

In this letter, by laser heating ionic ammonia above 1000 K, we report the formation of a previously unobserved ionic phase. Raman spectroscopy reveals a pronounced structural change, accompanied by the emergence of two distinct asymmetric H$_2$ vibron ($\nu_1$ and $\nu_2$). Comparison with first-principles Raman calculations identifies the newly formed phase as the theoretically predicted NH$_7$ structure in NH$_3$-H$_2$ mixture. This structure exhibits an unusual ionic framework composed of three distinct structural units: H$_2$ molecules, H$^-$, and NH$_4^+$ ions. Our results provide the first experimental evidence that a stable ammonia hydride can form under extremes conditions.

\begin{figure}[t]
	\centering
    \includegraphics[width=\columnwidth]{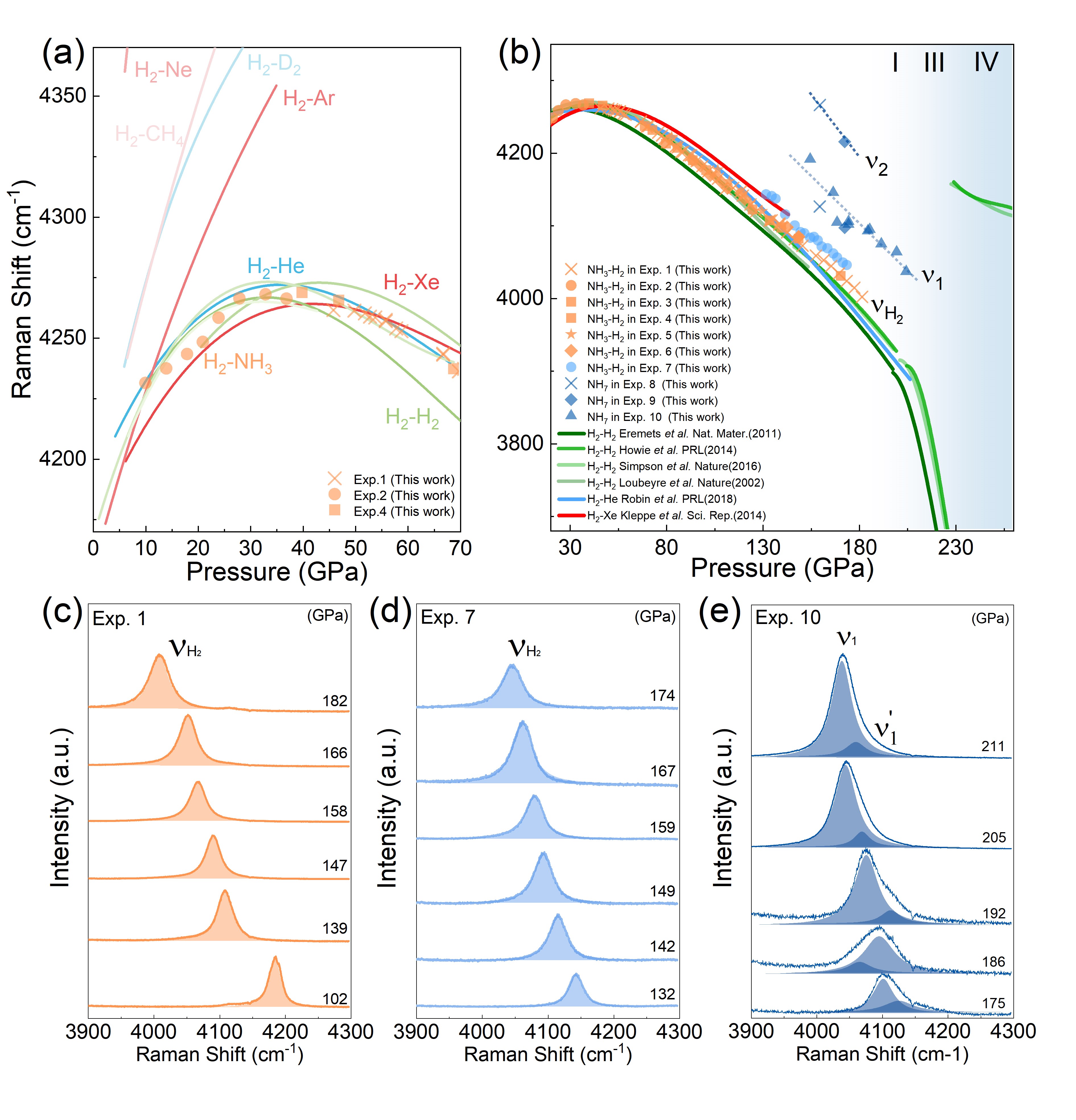}
    \caption{(a,b) Comparison of the pressure dependence of the H$_2$ vibron in different H$_2$-containing mixtures. Solid lines represent data taken from the Ref.\cite{2018_PRL_Turnbull,1992_PRB_Loubeyre,2014_SR_Kleppe,2014_PRL_Howie,2011_NM_Eremets,2016_Nature_Simpson,2002_Nature_Loubeyre}. Solid circles represent the H$_2$ vibron frequencies measured in the NH$_3$  in this work. Different symbols represent different experimental runs. Orange filled circles denote H$_2$ vibron ($v_{\mathrm{H}_2}$) observed in phase V ammonia without laser heating, forming a localized NH$_3$--H$_2$ mixture. Light bule filled circles represent H$_2$ vibron ($v_{\mathrm{H}_2}$) generated by laser heating phase V ammonia before the onset of ionic transition. Dark blue filled circles denote H$_2$ vibron generated by laser heating phase $\upbeta$ ionic ammonia. The $\nu_1$ and $\nu_2$ represent the two observed H$_2$ vibron in ionic NH$_7$ phase. (c--e) Representative Raman spectra of the H$_2$ vibrons under different conditions. $\nu_1^{\prime}$ denotes a weak shoulder on the asymmetric $\nu_1$ vibron mode (FIG. S2-S3). Other experiments are provided in the Supplemental Material.}
	\label{fig:fig2}
\end{figure}

\textit{Experimental Results}---At approximately 150 GPa, solid ammonia undergoes a phase transition from the molecular phase V to the ionic phase $\upbeta$. This phase transition is accompanied by pronounced changes in the Raman spectrum, together with the emergence of $\nu_b$ mode (FIG. 1). The existence and structure of this ionic phase were first established in earlier experimental studies and have recently been further confirmed by our previous work \cite{2014_PRB_Ninet,2014_NC_Palasyuk,2026_JPCL_Tao}. We performed laser heating after ammonia transformed into the phase$\upbeta$. No external laser absorber was used during the heating process. Upon laser heating, the initially transparent ionic ammonia transformed into an opaque phase (FIG. 1b inset, FIG S1). Raman spectra collected from the opaque region reveal six new lattice modes that are distinct from those of any previously reported ammonia phase. Several new Raman modes also emerge in the 1400--1800 cm$^{-1}$ region, which is general associated with N--H bending vibrations. These pronounced spectral changes provide clear evidence for a substantial structural change in ionic ammonia upon heating. Meanwhile, two distinct H$_2$ vibron modes ($v_1$ and $v_2$)with asymmetric shape line appear in the high-frequency region (Fig. 1,FIG. S2), indicating that laser heating also induces dissociation of ionic ammonia and the formation of molecular hydrogen. Indeed, we do not rule out the possibility of a reaction between NH$_3$ and the Re gasket. However, the characteristics of the newly observed Raman modes are inconsistent with those typically reported for Re--N compounds\cite{2023_IG_Zhang,2019_NC_Bykov}. 

It is clear that laser heating of the $\upbeta$ phase produces a new structure. To further confirmed whether this transformation involves a reaction between ionic ammonia and molecular hydrogen, we further compare the behavior of H$_2$ generated under different conditions in ammonia with that of pure solid hydrogen. We first tracked the behavior of hydrogen released by ammonia dissociation in phases V, at pressures well below the ionic transition (Exp. 1-6). Under these conditions, localized regions of an NH$_3$--H$_2$ mixture formed within the sample chamber. We tracked the H$_2$ vibron over a broad pressure range of 10--190 GPa. The pressure dependence of this vibron in NH$_3$--H$_2$ mixture closely follows that of pure H$_2$ throughout the entire range. For H$_2$-containing mixtures, a characteristic signature of van der Waals compound formation is a pronounced modification of the turnover behavior of the H$_2$ vibron (FIG. 2a). Notably, even in the H$_2$--D$_2$ system, where no van der Waals compound is formed, the pressure dependence of the H$_2$ vibron is still significantly modified \cite{1992_PRB_Loubeyre,2014_PRL_Howie}. Therefore, once molecular hydrogen interacts appreciably with other molecular species, its vibron behavior is expected to be noticeably modified. Moreover, even after molecular ammonia transforms into the ionic phase $\upbeta$, the vibron of H$_2$ in ionic ammonia remains essentially unchanged, indicating that H$_2$ does not react with the $\upbeta$ phase under cold compression (FIG.2b, FIG. S6). We further performed laser heating before the onset of the ionic transition (Exp. 7), which generated a clear H$_2$ vibron signal due to the thermal decomposition. Upon subsequent compression, the pressure evolution of this thermally produced hydrogen remains closely consistent with that of pure H$_2$. Meanwhile, no formation of new phases was observed after laser heating in this experiment. These H$_2$ vibron modes ($v_{\mathrm{H}_2}$) exhibit well-defined symmetric line shapes. This is consistent with previous reports of laser-heated phase V at lower pressures \cite{2012_JCP_Ojwang,2019_PRB_Ninet}. Hence, prior to the formation of the $\upbeta$ phase, neither cold compression nor laser heating leads to detectable chemical interaction between NH$_3$ and H$_2$ or to the formation of a new NH$_3$--H$_2$ compound.

In contrast, hydrogen generated after the ionization of ammonia exhibits markedly different spectroscopic behavior. In two experiments involving laser heating at 160 and 173 GPa, respectively, two distinct H$_2$ vibron modes, denoted $\nu_1$ and $\nu_2$, were observed (FIG. 1b). We performed several laser-heating experiments on ionic ammonia phase $\upbeta$. In most experiments, catastrophic diamond failure occurred during laser heating, whereas in the above two successful experiments the diamonds also failed after the Raman signals of the transformed sample had been collected. The markedly reduced stability of the diamond anvils during heating suggests that ionic ammonia undergoes substantially more rapid decomposition and/or chemical reactions at high temperature than molecular ammonia. Both modes exhibit pronounced asymmetric line shapes (FIG. S2-S3). In another experiment performed at 200 GPa, only a single H$_2$ vibron was resolved ($v_1$). Despite its single-peak character, its frequency lies substantially above that of pure H$_2$ at the same pressure, and the peak itself also displays a certain asymmetry. In contrast, the H$_2$ vibron ($v_{\mathrm{H}_2}$) peaks observed in the other experiments exhibit nearly symmetric line shapes (FIG. 2c--e). We rule out laser-heating-induced stress as the primary cause of this anomalously high frequency. In Exp. 7, a similar laser-heating procedure was employed. No anomalous H$_2$ vibron behavior was observed. Moreover, the lattice and N--H bending modes below 2000 cm$^{-1}$ in this experiment, exhibit pressure evolutions consistent with those observed in the other experimental (FIG. S10-S11). It suggest that the anomalous H$_2$ vibron behavior is intrinsic to the newly formed phase rather than a consequence of laser-heating-induced stress. Upon decompression, the frequency of this mode evolves continuously toward and eventually overlaps with the $\nu_1$ branch observed in the other two experiments. To further characterize $\nu_1$ and $\nu_2$ modes, we extracted their full widths at half maximum (FWHM) and compared them with those of pure phase-I hydrogen (FIG. S7). The linewidths are comparable to those of pure phase I over the corresponding pressure range. This similarity indicates that the observed vibron modes originate from molecular H$_2$ and are consistent with the rotationally disordered character of pure phase-I hydrogen.

The two emergences of H$_2$ vibron mode ($v_1$ and $v_2$) that deviate markedly from those of pure hydrogen, together with the appearance of new lattice modes after laser heating, provides clear evidence for the formation of a new NH$_3$--H$_2$ compound. For the NH$_3$--H$_2$ system, the high hydrogen content makes direct structural determination by synchrotron X-ray diffraction particularly challenging, due to the intrinsically weak X-ray X-ray scattering cross section. To identify this phase, we calculated the Raman spectra of candidate structures previously predicted for the NH$_3$--H$_2$ system and compared them with our experimental Raman spectra. For the NH$_3$--H$_2$ system, theoretical studies have mainly predicted the existence of $R\bar{3}m$- and $P4_12_12$-NH$_7$, as well as $Pm$-NH$_{10}$ \cite{2019_JPCL_Song,2024_PRM_He}. Among the proposed structures, the Raman spectrum of the $P4_12_12$-NH$_7$ phase shows good agreement with that of the new phase recovered to room temperature. On the other hand, phonon calculations indicate that another $R\bar{3}m$ candidate structure is dynamically unstable at 150 GPa (FIG. S12). We further calculated the pressure dependence of the Raman-active modes of $P4_12_12$-NH$_7$. The evolution of the lattice modes with pressure agrees closely with experiment, providing an additional constraint on the structural assignment. More importantly, the calculation also confirmed two distinct H$_2$ vibron modes, and both their frequencies and the separation between the two vibron branches are consistent with the experimentally observed $\nu_1$ and $\nu_2$ modes (FIG. 3a). In addition to the two dominant vibron modes, the calculations predict two much weaker Raman-active modes located on either side of the main peaks. These weak neighboring modes naturally account for the asymmetric line shapes observed experimentally after laser heating (FIG. S8). It is noteworthy that the theoretically predicted transition between the $R\bar{3}m$ and $P4_12_12$ structures occurs at approximately 60 GPa. In our experiments, however, the diamond anvils failed during decompression before the pressure could be reduced below $\sim$120 GPa. By comparison, substantially lower pressures can typically be reached during decompression in polymeric-nitrogen experiments. We speculate that the reduced diamond stability may be related to hydrogen generated during laser heating and its subsequent diffusion into the diamond anvils. Consequently, we were unable to determine whether the synthesized $P4_12_12$-NH$_7$ phase undergoes a structural transition in to $R\bar{3}m$ phase upon further decompression to lower pressures.

\begin{figure}[t]
	\centering
	\includegraphics[width=\columnwidth]{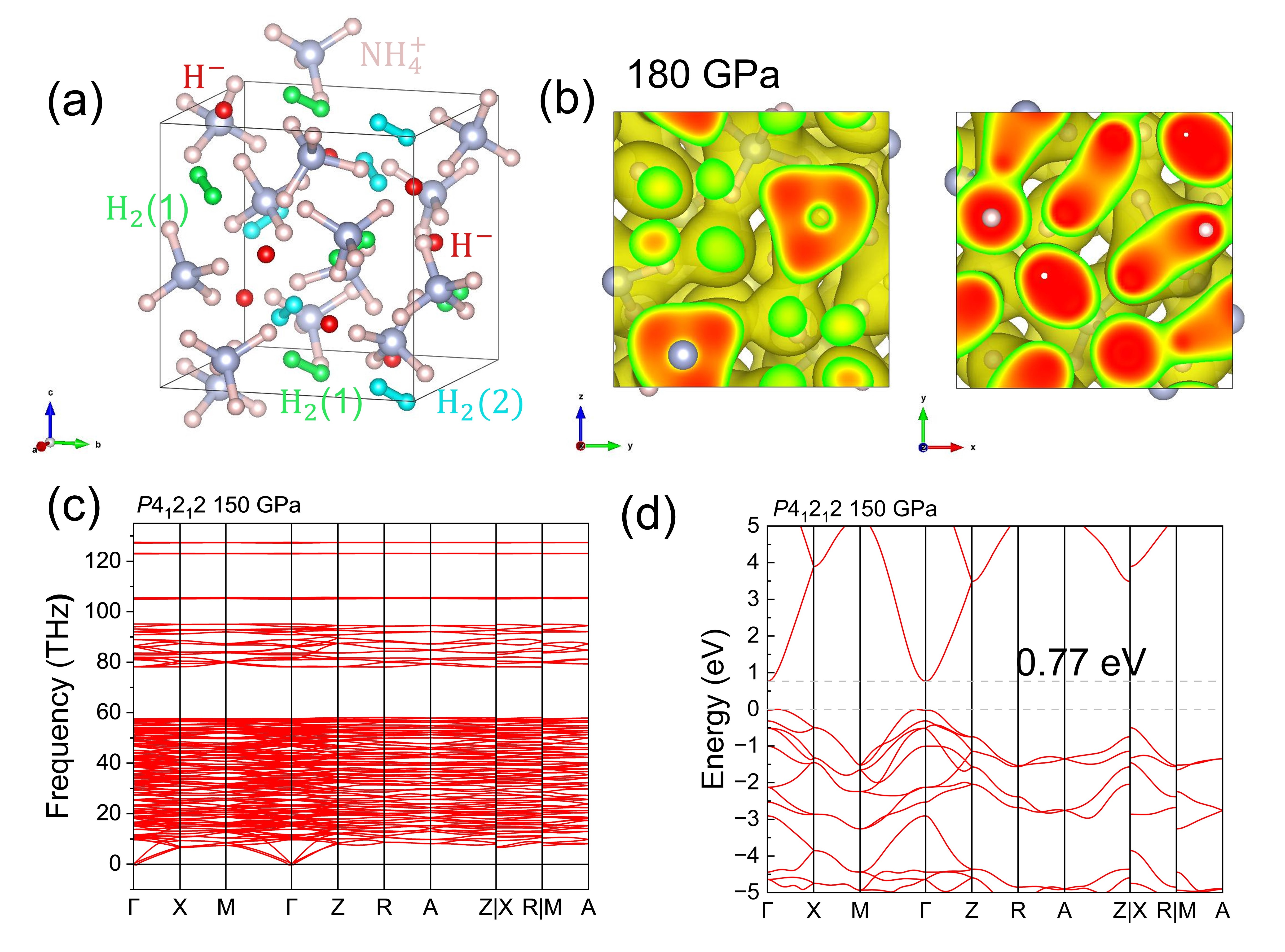}
	\caption{Structural and electronic properties of ionic NH$_7$. (a) Crystal structure of ionic NH$_7$. Hydrogen atoms belonging to different structural units, including H$_2$ molecules, NH$_4^+$ ions, and H$^-$ species, are shown in different colors. The H$_2$ molecules occupy two distinct local environments. (b) Electron localization function (ELF) isosurfaces at ELF = 0.5 viewed along different crystallographic directions at 180 GPa. (c) Calculated phonon dispersion of ionic NH$_7$ at 150 GPa. (d) Electronic band structure calculated at 150 GPa using PAW pseudopotentials.}
	\label{fig:fig3}
\end{figure}

The $P4_12_12$-NH$_7$ structure can be described as consisting of three distinct units: NH$_4^+$, molecular H$_2$, and H$^-$ species. This structure was originally predicted to form in the NH$_3$--H$_2$ system above approximately 50 GPa under cold-compression conditions \cite{2019_JPCL_Song}. Our experiments, however, show that NH$_3$ and H$_2$ remain chemically unreactive under cold compression up to at least 200 GPa. Formation of this structure requires further laser heating after ammonia undergoes the ionic phase transition. The formation of NH$_7$ therefore appears to be kinetically hindered and requires thermal activation to overcome a substantial reaction barrier. Such a requirement for a reaction between phase $\upbeta$ and H$_2$ is consistent with both our observations and previous studies that NH$_7$ was not formed when molecular phase V was laser heated. These results suggest a formation pathway in which laser heating induces partial dissociation of ammonia and generates molecular H$_2$, which subsequently reacts with phase $\upbeta$ and undergoes structural reorganization in hot dense ammonia to form $P4_12_12$-NH$_7$. Although ammonia decomposition can also produce molecular nitrogen, the laser-heating temperatures in our experiments remained below 2000 K, and no evidence for the formation of polymeric nitrogen was observed \cite{2010_PRL_Laniel,2020_SA_Ji,2004_NM_Eremets}. This path could also explain why NH$_7$ was not observed in previous experiments in which molecular ammonia was laser heated to temperatures exceeding 2000 K\cite{2012_JCP_Ojwang,2019_PRB_Ninet}.

Notably, the NH$_7$ structure contains H$_2$ molecules in two distinct local environments. To examine the bonding characteristics of this unusual ionic framework, we calculated the electron localization function (ELF). Up to 180 GPa, the H$_2$ units remain well defined and show only weak electronic interaction with the surrounding ionic ammonia framework [Fig. 3(b)]. This weak interaction provides a qualitative explanation for the absence of a resolved H$_2$-vibron doublet. During recovery to room temperature, the molecular H$_2$ units could be  released from the NH$_7$ framework. Previous theoretical studies also suggested that both NH$_7$ and NH$_{10}$ serve as potential hydrogen-storage materials \cite{2024_PRM_He}.

\begin{figure}[t]
	\centering
	\includegraphics[width=\columnwidth]{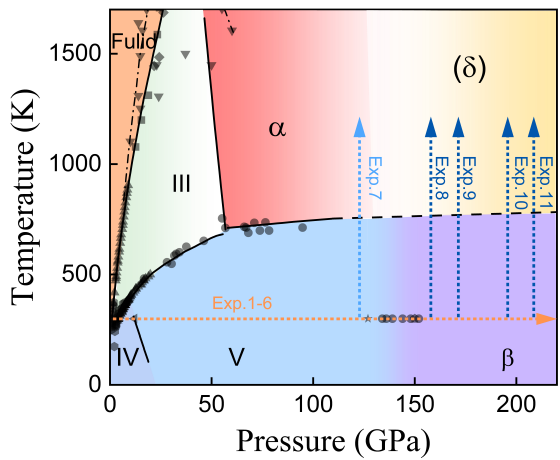}
	\caption{High-pressure and high-temperature phase diagram of ammonia. Solid lines and filled symbols indicate experimentally determined phase boundaries from Ref.\cite{2014_PRB_Ninet,2012_JCP_Ojwang,2012_PRL_Ninet,2014_NC_Palasyuk,1980_JPC_Hanson,2026_JPCL_Tao,2009_PRB_Datchi}. The dotted line denotes the melting curve of ammonia \cite{2019_PRB_Ninet,2012_JCP_Ojwang}. Dashed arrows indicate the paths explored in this work. Dark blue arrows correspond to experiments in which the ionic NH$_7$ phase was observed after heating. Light blue dashed arrows denote experiments in which no NH$_7$ phase was detected. AIMD simulations indicate that ionic $P4_12_12$-NH$_7$ can transform into a superionic state at elevated temperatures \cite{2019_JPCL_Song}. An additional superionic phase ($\updelta$) may exist above phase $\upbeta$.The black dashed line denotes the estimated $\upbeta$--$\updelta$ phase boundary extrapolated from the experimentally determined $\upalpha$--V phase boundary.}
	\label{fig:fig4}
\end{figure}

The pronounced change in sample transparency upon laser heating indicates a substantial modification of the electronic band structure. It suggests a possible transition from an insulating to a semiconducting state. Hence, we further calculated the electronic band structures of the ionic NH$_7$ phase and the ionic ammonia before laser heating. $Pma2$ and $Pca2_{1}$ are currently regarded as the two leading structural candidates for ionic ammonia \cite{2014_NC_Palasyuk,2014_PRB_Ninet}. Our calculations show that both structures are insulating, with band gaps of approximately 3.2 and 4.5 eV, respectively [FIG. S14]. In contrast, $P4_12_12$-NH$_7$ exhibits a much narrower band gap of only $\sim$0.7 eV at the corresponding pressure. Combined with previous calculations at lower pressures, the band gap of $P4_12_12$-NH$_7$ shows an approximately linear decrease with increasing pressure. This approximately linear trend is consistent with experimental observations reported for molecular crystals\cite{2022_PRL_Loubeyre}. The pronounced narrowing of the electronic gap also provides a natural explanation for the experimentally observed transformation from transparent ionic ammonia to an opaque phase after laser heating. Notably, the candidate $Pm$-NH$_{10}$ structure is predicted to remain insulating, which is difficult to reconcile with the pronounced optical darkening observed experimentally.

The transformation from ionic ammonia to ionic NH$_7$ is irreversible upon recovery to room temperature. This behavior suggests that, beyond the stability field of the $\upbeta$ phase, an additional high-temperature phase may become accessible. The previously established superionic $\upalpha$ phase of ammonia is produced by laser heating the molecular V phase. Above approximately 57 GPa, heating phase V to temperatures above 700 K drives ammonia into a superionic state. Upon recovery to room temperature, however, the sample reverts to phase V \cite{2014_PRB_Ninet,2026_PRB_Guigue}. In contrast, the $P4_12_12$-NH$_7$ structure identified here contains an intrinsically ionic hydrogen sublattice with protonic species, a structural feature that may strongly favor proton mobility at elevated temperatures. Previous AIMD simulations support the transformation of ionic NH$_7$ into a superionic state at elevated temperatures \cite{2019_JPCL_Song}. We therefore propose that, above approximately 150 GPa and 800 K, an additional superionic regime, denoted the $\updelta$ phase, could exist in the high-pressure ammonia system (FIG. 4).

In conclusion, by laser heating ionic ammonia above 1000 K, we observed the formation of a previously unreported ionic ammonia hydride. Raman spectra and first-principles calculations identify this phase as the theoretically predicted ionic $P4_12_12$-NH$_7$. Ionic ammonia undergoes partial dissociation and pronounced structural change, with hydrogen reoranization ultimately leading to the formation of $P4_12_12$-NH$_7$. Our results reveal that the formation of ammonia hydrides follows a thermodynamic pathway distinct from those of other NH$_2$-containing mixtures. The exotic hydrogen bonding in ionic NH$_7$ suggests the possible existence of a new superionic phase ($\updelta$) at high temperature of ionic ammonia. 

	\textit{Acknowledgments}---This work was financially supported by the National Natural Science Foundation of China (Grant No. 12374013) and the Fundamental Research Funds for the Central University (Grant No. 2020SCUNL107).
	
   	\textit{Author contributions}---Li Lei conceived and supervised the project. Jingyi Liu, Yu Tao and Li Lei conducted the  experiments. Yu Tao performed the theoretical calculations, analyzed the data and prepared the manuscript. Li Lei revised the manuscript.
	
	\bibliographystyle{apsrev4-2}
	\bibliography{refs}
	
\end{document}